\documentclass[11pt, a4paper]{article}
\usepackage{fullpage}

\usepackage[T1]{fontenc}
\usepackage[utf8]{inputenc}
\usepackage{csquotes}
\usepackage[english]{babel}
\usepackage{tipa}

\usepackage{xcolor}

\usepackage{amsmath}
\usepackage[overload]{empheq}
\usepackage{authblk}

\DeclareMathOperator*{\avg}{avg}

\usepackage{hyperref}
\hypersetup{
	colorlinks,
	linktoc=page,
	linkcolor=linkcolor,
	citecolor=linkcolor,
	urlcolor=linkcolor
}

\usepackage{acronym}
\usepackage{listings}

\def\includegraphics{}

\definecolor{lifex}{HTML}{f60248}
\newcommand{\lifex}{\texttt{life\textsuperscript{\color{lifex}{x}}}}

\definecolor{codegreen}{rgb}{0,0.6,0}
\definecolor{codegray}{rgb}{0.5,0.5,0.5}
\definecolor{codepurple}{rgb}{0.58,0,0.82}
\definecolor{backcolor}{rgb}{0.95,0.95,0.92}
\colorlet{linkcolor}{blue!50!black}

\lstdefinestyle{mystyle}{
    backgroundcolor=\color{backcolor},
    commentstyle=\color{codegreen},
    keywordstyle=\color{lifex},
    numberstyle=\tiny\color{codegray},
    stringstyle=\color{codepurple},
    basicstyle=\ttfamily\footnotesize,
    breakatwhitespace=false,
    breaklines=true,
    captionpos=b,
    keepspaces=true,
    numbers=none,
    numbersep=5pt,
    showspaces=false,
    showstringspaces=false,
    showtabs=false,
    tabsize=4,
    frame=shadowbox
}

\lstdefinelanguage{prm}{
    keywords={subsection, set, end},
    comment=[l]{\#}
}

\lstdefinelanguage{xml}{
    keywords={value, default\_value, documentation, pattern, pattern\_description}
}

\lstdefinelanguage{json}{
    keywords={value, default\_value, documentation, pattern, pattern\_description}
}

\begin{document}

%\begin{frontmatter}

%\begin{fmbox}
%\dochead{Software}

\title{An open tool based on \texorpdfstring{\lifex{}}{lifex} for myofibers generation in cardiac computational models}

\date{}

\author[a]{Pasquale C. Africa}
\author[a]{Roberto Piersanti}
\author[a]{Marco Fedele}
\author[a]{Luca Dede'}
\author[a,b]{Alfio Quarteroni}

\affil[a]{\footnotesize MOX, Department of Mathematics, Politecnico di Milano, Italy}
\affil[b]{\footnotesize Institute of Mathematics, \'Ecole Polytechnique Fédérale de Lausanne, Switzerland (Professor emeritus)}

\maketitle

%\author[addressref={mox}, email={pasqualeclaudio.africa@polimi.it}]{\inits{P.C.A.}\fnm{Pasquale Claudio} \snm{Africa}}
%\author[addressref={mox}]{\inits{R.P.}\fnm{Roberto} \snm{Piersanti}}
%\author[addressref={mox}]{\inits{M.F.}\fnm{Marco} \snm{Fedele}}
%\author[addressref={mox}]{\inits{L.D.}\fnm{Luca} \snm{Dede'}}
%\author[addressref={mox,epfl}]{\inits{A.Q.}\fnm{Alfio} \snm{Quarteroni}}
%
%\address[id=mox]{
%  \orgdiv{MOX, Department of Mathematics},
%  \orgname{Politecnico di Milano},
%  \city{Milano},
%  \cny{Italy}
%}
%\address[id=epfl]{
%  \orgdiv{Institute of Mathematics},
%  \orgname{\'Ecole Polytechnique Fédérale de Lausanne},
%  \city{Lausanne},
%  \cny{Switzerland (Professor emeritus)}
%}

%\end{fmbox}

%\begin{abstractbox}

\begin{abstract}
\textbf{Background}:
Modeling the whole cardiac function involves the solution of several complex multi-physics and multi-scale models that are highly computationally demanding, which call for simpler yet accurate, high-performance computational tools. Despite the efforts made by several research groups, no software for whole-heart fully-coupled cardiac simulations in the scientific community has reached full maturity yet.

\textbf{Results}:
In this work we present the first publicly released package of \lifex{}, a high-performance \acl{FE} solver for multi-physics and multi-scale problems developed in the framework of the iHEART project. The goal of \lifex{} is twofold. On the one side, it aims at making \textit{in silico} experiments easily reproducible and accessible to a wide community of users, including those with a background in medicine or bio-engineering. On the other hand, as an academic research library \lifex{} can be exploited by scientific computing experts to explore new mathematical models and numerical methods within a robust development framework. \lifex{} has been developed with a modular structure and will be released bundled in different modules. The tool presented here proposes an innovative generator for myocardial fibers based on \aclp{LDRBM}, which are the essential building blocks for modeling the electrophysiological, mechanical and electromechanical cardiac function, from single-chamber to whole-heart simulations.

\textbf{Conclusions}:
The tool presented in this document is intended to provide the scientific community with a computational tool that incorporates general state of the art models and solvers for simulating the cardiac function within a high-performance framework that exposes a user- and developer-friendly interface. This report comes with an extensive technical and mathematical documentation to welcome new users to the core structure of a prototypical \lifex{} application and to provide them with a possible approach to include the generated cardiac fibers into more sophisticated computational pipelines. In the near future, more modules will be successively published either as pre-compiled binaries for \texttt{x86-64 Linux} systems or as open source software.
\end{abstract}

%\begin{keyword}
%    \kwd{computational cardiology}
%    \kwd{high-performance computing}
%    \kwd{cardiac fibers}
%    \kwd{mathematical modeling}
%    \kwd{finite element methods}
%\end{keyword}

%\begin{keyword}[class=AMS]
%    \kwd[Primary ]{68-04, 68N30}
%    \kwd[; secondary ]{35-04, 65-04, 65M60, 65N30, 65Y05, 65Y20, 92-04, 92C50}
%\end{keyword}

%\end{abstractbox}

%\end{frontmatter}

\section*{Background}\label{sec:intro}
The human heart function is a complex system involving interacting processes at the molecular, cellular, tissue, and organ levels with widely varying time scales. For this reason, it is still among the most arduous modeling and computational challenges in a field where \textit{in silico} models and experiments are essential to reproduce both physiological and pathological behaviors \cite{book_cardiovascular}.

A satisfactorily accurate model for the whole cardiac function must be able to describe a wide range of different processes, such as: the propagation of the trans-membrane potential and the flow of ionic species in the myocardium, the deformation caused by the muscle contraction, the dynamics of the blood flow through the heart chambers and cardiac valves \cite{iheart2017}. In particular, the dynamics of ionic species needs for accurate models specifically designed to reproduce physiological \cite{piersanti2021modeling} and pathological scenarios \cite{SALVADOR2021104674,salvador2022role} (such as the \textit{ten Tusscher--Panfilov} \cite{ttp06} and the \textit{Courtemanche-Ramirez-Nattel} \cite{crn} ionic models for ventricular/atrial cells, respectively).

These demanding aspects make whole-heart fully-coupled simulations computationally intensive and call for simpler yet accurate, high-performance computational tools.

\begin{figure}
	\centering
	\includegraphics[width=0.5\textwidth]{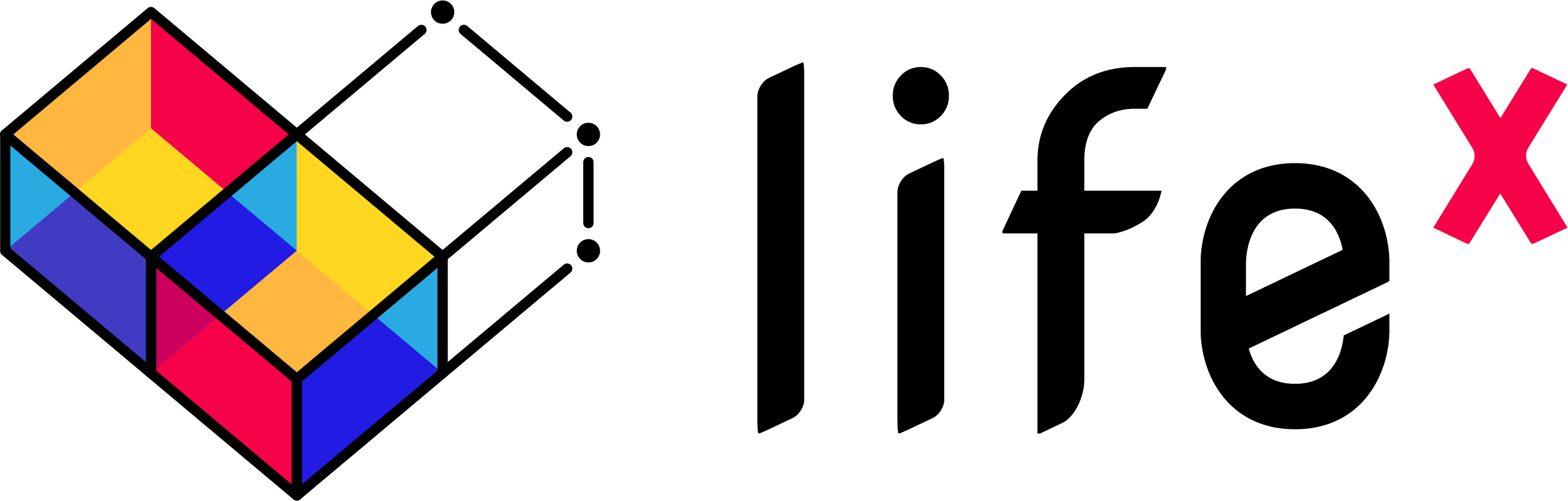}
	\caption{\lifex{} official logo.}
	\label{fig:logo}
\end{figure}

In this work we introduce \lifex{} (pronounced \textipa{/,la\textsci f\textprimstress\textepsilon ks/}), a high-performance \ac{FE} numerical solver for multi-physics and multi-scale differential problems. It is written in \texttt{C++} using the most modern programming techniques available in the \texttt{C++17} standard and is built upon the \texttt{deal.II}\footnote{\url{https://www.dealii.org/}} \cite{dealII93} \ac{FE} core. The official \lifex{} logo is shown in Figure~\ref{fig:logo}.

The code is natively parallel and designed to run on diverse architectures, ranging from laptop computers to \ac{HPC} facilities and cloud platforms. We tested our software on a cluster node endowed with 192 cores based on Intel Xeon Gold 6238R, 2.20 GHz, available at MOX, Dipartimento di Matematica, Politecnico di Milano, and on the \texttt{GALILEO100} supercomputer available at \texttt{CINECA} (Intel CascadeLake 8260, 2.40GHz, see \url{https://wiki.u-gov.it/confluence/display/SCAIUS/UG3.3%3A+GALILEO100+UserGuide} for more technical specifications).

Despite being conceived as an academic research library in the framework of the iHEART project (see Section~\nameref{sec:funding}), \lifex{} is intended to provide the scientific community with a \acl{FE} solver for real world applications that boosts the user and developer experience without sacrificing its computational efficiency and universality.

The initial development of \lifex{} has been oriented towards a \texttt{heart} module incorporating several state-of-the-art core models for the simulation of cardiac electrophysiology, mechanics, electromechanics, blood fluid dynamics and myocardial perfusion. Such models have been recently exploited for a variety of standalone or coupled simulations both under physiological and pathological conditions (see, \textit{e.g.}, \cite{regazzoni2022cardiac,SALVADOR2021104674,REGAZZONI2021104641,zingaro2022multiscale,bucelli2021,salvador2022role,fumagalli2021,STELLA2020104047,dede2021modeling,piersanti20213d0d}).

Since this first release is focused on modeling the cardiac fibers, in the next two paragraphs we will briefly review the physiology of myofibers and the most used mathematical methods to model them, describing in detail their implementation which is included within this \lifex{} version.

\subsection*{Cardiac fibers: physiology and modeling}
The heart is a four chambers muscular organ whose function is to pump the blood throughout the whole circulatory system. The upper chambers, the right and left atria, receive incoming blood. The lower chambers, the right and left ventricles, pump blood out of the heart and are more muscular than atria. The left heart (\textit{i.e.} left atrium and left ventricle) pumps the oxygenated blood through the systemic circulation, meanwhile the right heart (\textit{i.e.} right atrium and right ventricle) recycles the deoxygenated blood through the pulmonary circulation~\cite{quarteroni2017integrated,book_cardiovascular}. The atria and the ventricles are separated by the atrioventricular valves (mitral and tricuspid valves) that regulate the blood transfer from the upper to lower cavities. The four chambers are connected to the circulatory system: the ventricles with the aorta through the aortic valve and pulmonary artery via the pulmonary valve; the left atrium with the left and right pulmonary veins, whereas the right atrium with superior and inferior caval veins~\cite{quarteroni2017integrated,book_cardiovascular}.

The heart wall is made up of three layers: the internal thin \textit{endocardium}, the external thin \textit{epicardium} and the thick muscular cardiac tissue, the \textit{myocardium}. Most of the myocardium is occupied by \textit{cardiomyocytes}, striated excitable muscle cells that are joined together in linear arrays. The result of cluster cardiomyocytes, locally organized as composite laminar sheets, defines the orientation of muscular \textit{fibers} (also called \textit{myofibers}). Aggregations of myofibers give rise to the fiber-reinforced heart structure defining the cardiac muscular architecture~\cite{streeter1969fiber,legrice1995laminar}.

\begin{figure}
	\centering
	\includegraphics[width=\textwidth]{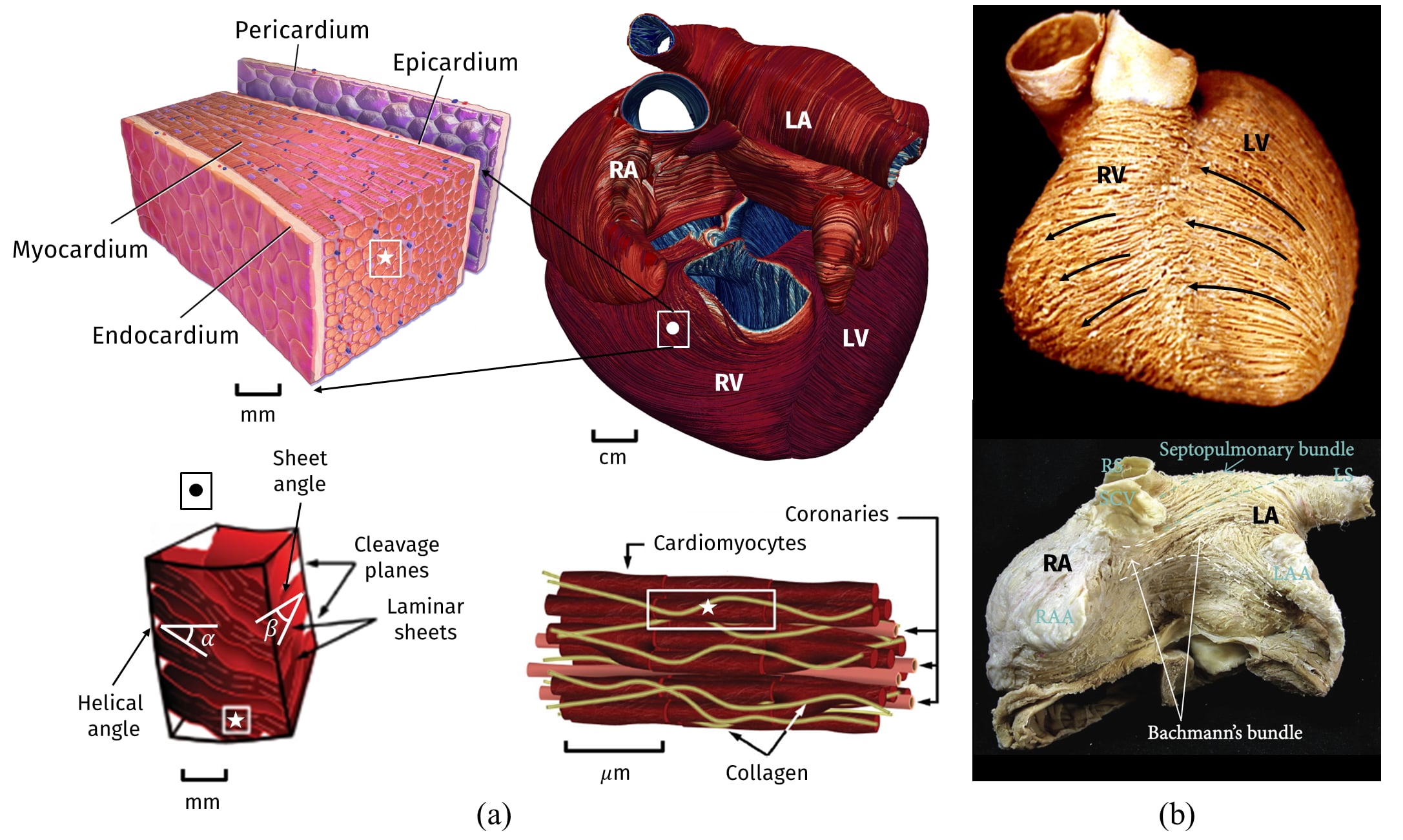}
	\caption{(a) Representation of the multiscale cardiac muscle. (b) Anatomical dissection of myocardial fibers in ventricles (top) and atria (bottom). Images taken and readapted from~\cite{chabiniok2016multiphysics,anderson2019cardiomyocytes,sanchez2014left,figure_heart_structure}. Images were available either freely under a Creative Commons Attribution license or have been granted reuse permission by the copyright holder.}
	\label{fig:fibers_anatomy}
\end{figure}

A schematic representation of the multiscale myocardial fiber-structure is shown in Figure~\ref{fig:fibers_anatomy}. Ventricular muscular fibers are well-organized as two intertwined spirals wrapping the heart around, defining the characteristic myocardial helical structure~\cite{greenbaum1981left,lombaert2012human}. Local orientation of myofibers is identified by their angle on the tangent plane and on the normal plane of the heart, called the \textit{helical} and the \textit{sheet} angles, respectively~\cite{streeter1969fiber,toussaint2013vivo}. The transition inside the myocardial wall is characterized by a continuous, almost linear change in helical angle from about $60^\circ$ at the epicardium to nearly $-60^\circ$ at the endocardium~\cite{lombaert2012human,streeter1969fiber}.

Atrial fibers architecture is very different from that of the ventricles, where myofibers are aligned in a regular pattern~\cite{streeter1969fiber}. Indeed, myofibers in the atria are arranged in individual bundles running along different directions throughout the wall chambers~\cite{ho2002atrial,ho2009importance}. Preferred orientation of myofibers in the human atria is characterized by multiple overlapping structures, which promote the formation of separate attached bundles~\cite{sanchez2013standardized}, as shown in Figure~\ref{fig:fibers_anatomy}. 

The cardiac muscular fiber architecture is the backbone of a proper pumping function and has a strong influence on the electric signal propagation throughout the myocardium and also on the mechanical contraction of the muscle~\cite{roberts1979influence,punske2005effect,eriksson2013influence,palit2015computational}.
This motivates the need to accurately include fibers orientation in cardiac computational models in order to obtain physically sound results~\cite{guan2020effect,piersanti2021modeling,piersanti20213d0d}. 

Due to the difficulty of reconstructing cardiac fibers from medical imaging, different methodologies have been proposed to provide a realistic surrogate of myofibers orientation~\cite{rossi2014thermodynamically,lombaert2012human,bayer2012novel,doste2019rule,hoermann2019automatic,wong2014generating,krueger2011modeling,ferrer2015detailed,fastl2018personalized,roney2020constructing}. Among these, atlas-based methods map and project a detailed fiber field, previously reconstructed on an atlas, on the geometry of interest, exploiting imaging or histological data~\cite{lombaert2012human,hoermann2019automatic,roney2020constructing}. However, these methods require complex registration algorithms and their results depend on the original atlas data upon which they have been built.

Alternative strategies for generating myofiber orientations are the \acp{RBM}~\cite{beyar1984computer,potse2006comparison,nielsen1991mathematical,wong2014generating,krueger2011modeling,ferrer2015detailed,bayer2012novel,piersanti2021modeling}. \Acp{RBM} describe fiber orientations with mathematically sound rules based on histological and imaging observations and only require information about the myocardial geometry~\cite{streeter1969fiber}. These methods parametrize the transmural and apico-basal directions in the entire myocardium in order to assign orthotropic (longitudinal, transversal and normal) myofibers~\cite{piersanti2021modeling}.

A particular class of \Acp{RBM}, which relies on the solution of Laplace boundary-value problems, is known as \acp{LDRBM}, addressed in~\cite{rossi2014thermodynamically,bayer2012novel,doste2019rule}
and recently analyzed under a unified mathematical formulation~\cite{piersanti2021modeling}. \acp{LDRBM} define the transmural and apico-basal directions by taking the gradient of harmonic functions (the potentials) corresponding to suitable Dirichlet boundary conditions. These directions are then properly rotated to match histological observations~\cite{streeter1969fiber,lombaert2012human,greenbaum1981left,ho2002atrial}.

This initial release includes a generator for myocardial fibers based on \acp{LDRBM} \cite{piersanti2021modeling}, with application to a number of different prototypical and realistic geometries (slab models, left ventricles and left atria).

\subsection*{\aclp{LDRBM}}\label{sec:methods}
In this section, we briefly recall the \acp{LDRBM} that stand behind the myocardial fiber generation. In Section~\nameref{sec:results} we will present several examples where we elaborate on how to reproduce and run the algorithms presented hereafter.

This \textit{getting started} guide presents \acp{LDRBM} for (ventricular and spherical) slabs, (based and complete) left ventricular and left atrial geometries. For further details about the \acp{LDRBM} presented here see also \cite{piersanti2021modeling}.

\begin{figure}
	\centering
	\includegraphics[width=1\textwidth]{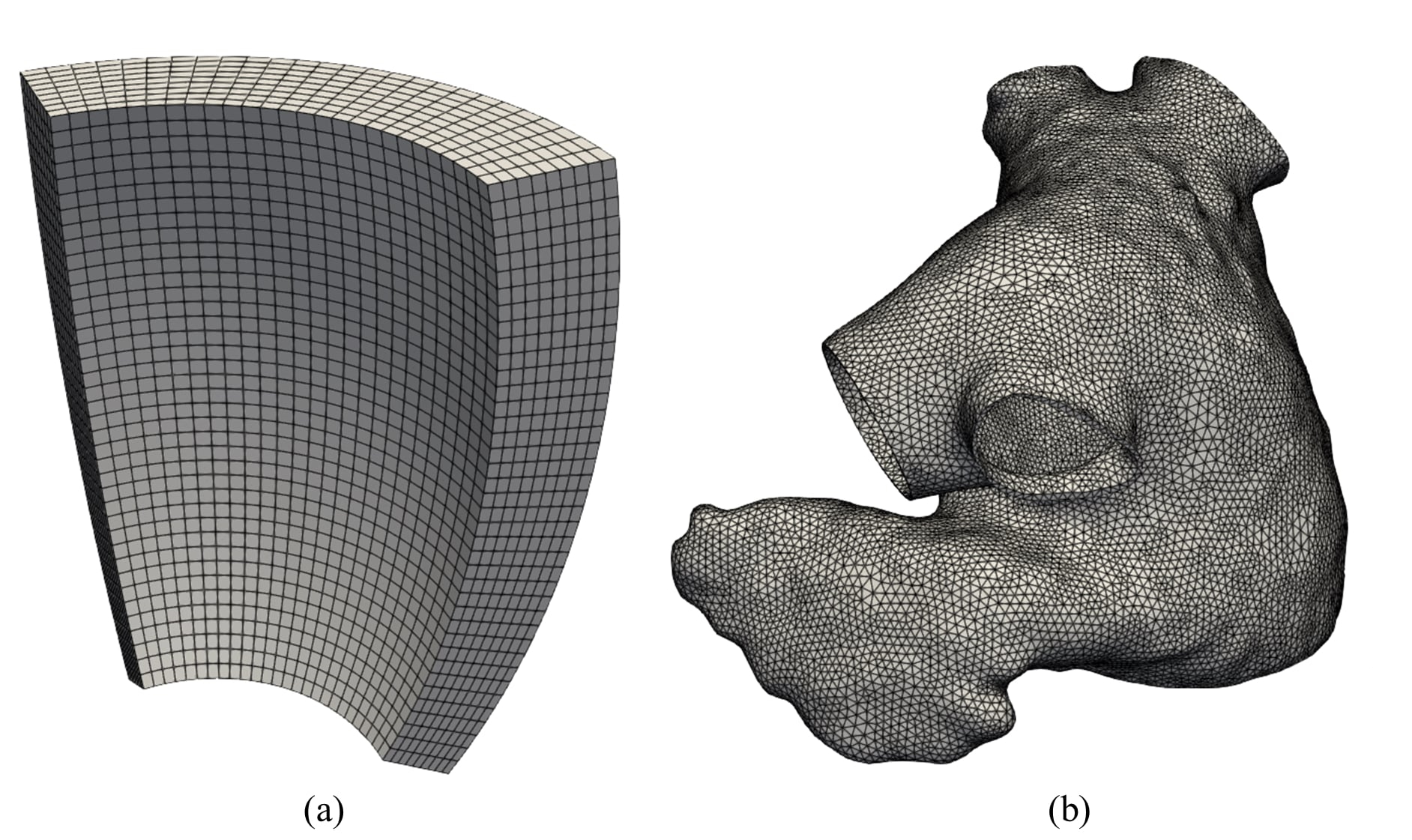}
	\caption{(a) Hexahedral mesh of a ventricular slab. (b) Tetrahedral mesh of a realistic left atrium \cite{fastl2018personalized}.}
	\label{fig:mesh}
\end{figure}

\begin{figure}
	\centering
	\includegraphics[width=1\textwidth]{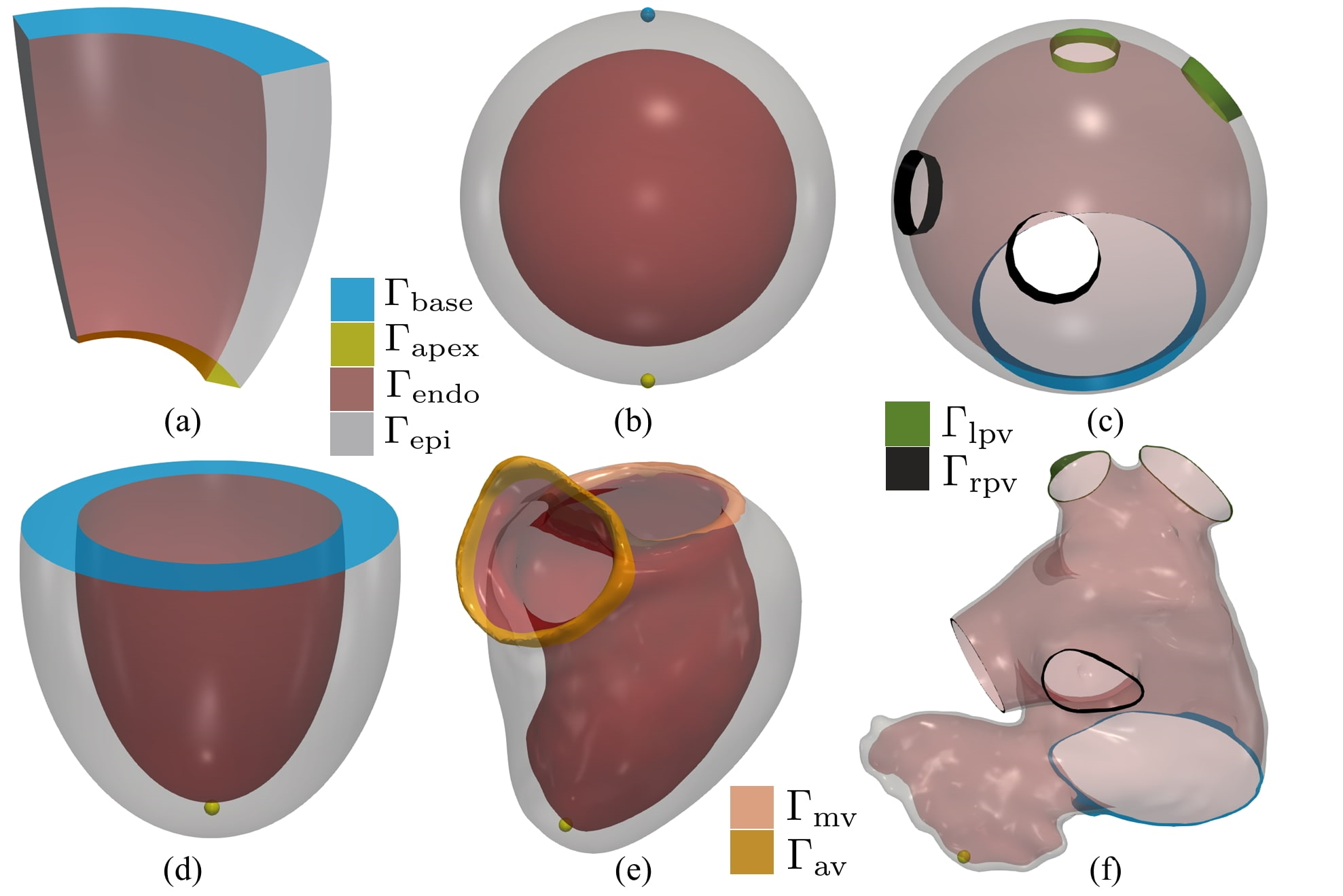}
	\caption{Labelled meshes. (a) Ventricular slab. (b) Spherical slab. (c) Idealized left atrium. (d) Idealized based left ventricle. (e) Realistic complete left ventricle. (f) Realistic left atrium. $\Gamma_\mathrm{base}$ denotes the basal plane, $\Gamma_\mathrm{apex}$ the apex, $\Gamma_\mathrm{endo}$ the endocardium, $\Gamma_\mathrm{epi}$ the epicardium, $\Gamma_\mathrm{lpv}$, $\Gamma_\mathrm{rpv}$ the left (right) pulmonary veins, respectively.}
	\label{fig:tags}
\end{figure}

\begin{figure}
	\centering
	\includegraphics[width=1\textwidth]{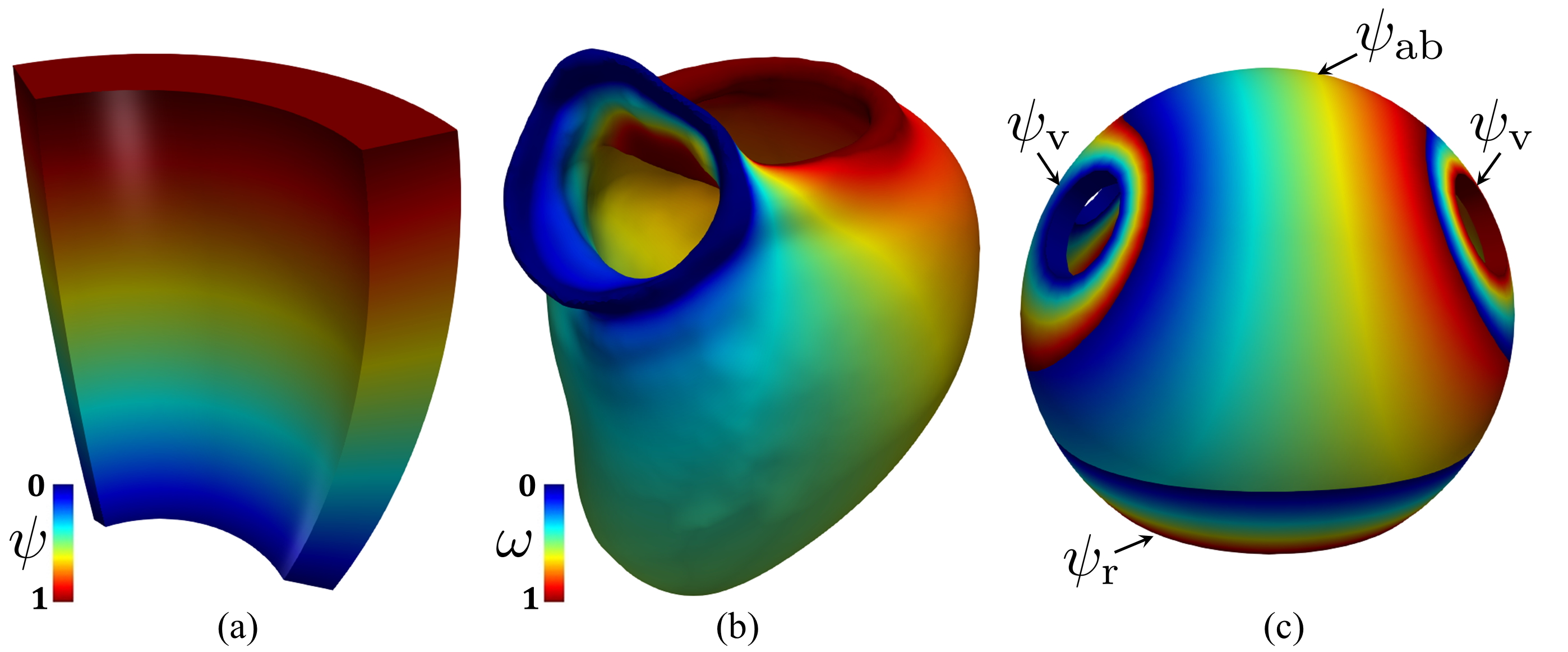}
	\caption{Different types of normal distances: (a) Bayer-Trayanova et al. approach \cite{bayer2012novel}. (b) Doste et al. approach \cite{doste2019rule}. (c) Piersanti et al. approach~\cite{piersanti2021modeling}.}
	\label{fig:laplace}
\end{figure}

The following common steps are the building blocks of all \acp{LDRBM}.
\begin{description}
        \item[1. Labelled mesh:] Provide a labelled volumetric mesh of the domain $\Omega$ to define specific partition of the boundary $\partial \Omega$ as
        $$
        \partial \Omega = \Gamma_\mathrm{epi} \cup \Gamma_\mathrm{endo} \cup \Gamma_\mathrm{base}
        \cup \Gamma_\mathrm{apex},
        $$

        where $\Gamma_\mathrm{endo}$ is the endocardium, $\Gamma_\mathrm{epi}$ is the
        epicardium, $\Gamma_\mathrm{base}$ is the basal plane and $\Gamma_\mathrm{apex}$ is the apex, which are demarcated through proper surface labels included in the input mesh. \lifex{} is designed to support both hexahedral and tetrahedral labelled meshes in the widely used \verb|*.msh| format, see Figure~\ref{fig:mesh}. This type of mesh can be generated by a variety of mesh generation software (\textit{e.g.} \texttt{gmsh}\footnote{\url{http://gmsh.info}}, \texttt{netgen}\footnote{\url{https://ngsolve.org/}}, \texttt{vmtk}\footnote{\url{http://www.vmtk.org}} and \texttt{meshtools}\footnote{\url{https://bitbucket.org/aneic/meshtool/src/master/}}). Otherwise, other mesh-format types can be converted in \verb|*.msh| using for example the open-source library \verb|meshio|\footnote{\url{https://pypi.org/project/meshio/}}.

        \begin{description}
                \item[Epicardium and endocardium:]
                for the ventricular slab geometry $\Gamma_\mathrm{endo}$ and $\Gamma_\mathrm{epi}$ are the lateral walls of the slab, see Figures~\ref{fig:tags}(a); for the spherical slab, the ventricular and atrial geometries $\Gamma_\mathrm{endo}$ and $\Gamma_\mathrm{epi}$ are the internal and external surfaces, see Figures~\ref{fig:tags}(b-f).

                \item[Basal plane and apex:] for the ventricular slab geometry $\Gamma_\mathrm{base}$ and $\Gamma_\mathrm{apex}$ are the top and bottom surfaces, respectively (see Figure~\ref{fig:tags}(a)); for the spherical slab, $\Gamma_\mathrm{base}$ and $\Gamma_\mathrm{apex}$ are selected as the north and south pole points of the epicardial sphere (see Figure~\ref{fig:tags}(b)); for the based ventricular geometry $\Gamma_\mathrm{base}$ is an artificial basal plane located well below the cardiac valves (see Figure~\ref{fig:tags}(d)), while for the complete ventricular geometry $\Gamma_\mathrm{base}$ is split into $\Gamma_\mathrm{mv}$ and $\Gamma_\mathrm{av}$, representing the mitral and aortic valve rings, respectively (see Figure~\ref{fig:tags}(e)); for the ventricular geometries $\Gamma_\mathrm{apex}$ is selected as the epicardial point furthest from the ventricular base (see Figures~\ref{fig:tags}(d-e)); for the atrial geometry $\Gamma_\mathrm{base}$ is the mitral valve ring and $\Gamma_\mathrm{apex}$ represents the apex of the left atrial appendage (see Figures~\ref{fig:tags}(c-f));

                \item[Atrial pulmonary rings:] the atrial geometry type also requires the definition of the boundary labels for the left $\Gamma_\mathrm{lpv}$ and right $\Gamma_\mathrm{rpv}$ pulmonary vein rings,  see Figures~\ref{fig:tags}(c-f).
        \end{description}

        \item[2. Transmural direction:] A transmural distance $\phi$ is defined to compute the distance of the epicardium from the endocardium, by
        means of the following \ac{LD} problem:
        \begin{equation}[left=\empheqlbrace]
        \begin{alignedat}{3}
        -\Delta \phi&=0, &\qquad&{\text{in }}\Omega,
        \\
        \phi &= 1, &\qquad&{\text{on }}\Gamma_\mathrm{epi},
        \\
        \phi &= 0, &\qquad&{\text{on }}\Gamma_\mathrm{endo},
        \\
        \nabla \phi \cdot \mathbf{n}&=0, &\qquad&{\text{on }}\partial \Omega\setminus(\Gamma_\mathrm{endo} \cup \Gamma_\mathrm{epi}).
        \end{alignedat}
        \end{equation}

         Then, the transmural distance gradient $\nabla \phi$ is used to build the unit transmural direction:
         $$
         \widehat{\boldsymbol{e}}_t=\frac{\nabla \phi}{\Vert \nabla \phi \Vert}.
         $$

         \item[3. Normal direction:] A normal (or apico-basal) direction $\boldsymbol{k}$ (which is directed from the apex towards the base) is introduced and used to build the unit normal direction $\widehat{\boldsymbol{e}}_n$:
         $$
         \widehat{\boldsymbol{e}}_n = \frac{\boldsymbol{k} - (\boldsymbol{k} \cdot
                \widehat{\boldsymbol{e}}_t )\widehat{\boldsymbol{e}}_t}{\Vert
                \boldsymbol{k} - (\boldsymbol{k} \cdot \widehat{\boldsymbol{e}}_t
                )\widehat{\boldsymbol{e}}_t \Vert}.
         $$

         The normal direction $\boldsymbol{k}$ can be computed following one of these approaches (see also Figure~\ref{fig:laplace}):
         \begin{description}
                \item[\ac{RL} et al. approach \cite{rossi2014thermodynamically}:] $\boldsymbol{k}$ is defined as the vector $\mathbf{n}_\mathrm{base}$, \textit{i.e.} the outward normal to the basal plane, that is $\boldsymbol{k}=\mathbf{n}_\mathrm{base}$.

                \item[\ac{BT} et al. approach \cite{bayer2012novel}:] $\boldsymbol{k}$ is the gradient of the solution $\psi$ ($\boldsymbol{k}=\nabla \psi$), which can be obtained by solving the following \ac{LD} problem:
                \begin{equation}[left=\empheqlbrace]
                \label{normal}
                \begin{alignedat}{3}
                -\Delta \psi&=0, &\qquad&{\text{in }}\Omega,
                \\
                \psi &= 1, &\qquad&{\text{on }}\Gamma_\mathrm{base},
                \\
                \psi &= 0, &\qquad&{\text{on }}\Gamma_\mathrm{apex},
                \\
                \nabla \psi \cdot \mathbf{n}&=0, &\qquad&{\text{on }}\partial \Omega
                \setminus(\Gamma_\mathrm{base} \cup \Gamma_\mathrm{apex}).
                \end{alignedat}
                \end{equation}

                \item[Doste et al. approach \cite{doste2019rule}:] $\boldsymbol{k}$ is a weighted sum of the apico-basal ($\nabla \psi_\mathrm{ab}$) and apico-outflow-tract ($\nabla \psi_\mathrm{ot}$)
                directions, obtained using an interpolation function $w$:
                $$
                \boldsymbol{k} = w\nabla \psi_\mathrm{ab} + (1-w)\nabla \psi_\mathrm{ot},
                $$
                where $\psi_\mathrm{ab}$ and $\psi_\mathrm{ot}$ are obtained by solving \ac{LD} problems in the form of \eqref{normal} where $\Gamma_\mathrm{base}=\Gamma_\mathrm{mv}$ (for $\psi_\mathrm{ab}$)
                and $\Gamma_\mathrm{base}=\Gamma_\mathrm{av}$ (for $\psi_\mathrm{ot}$), respectively. Moreover, the interpolation function $w$ is obtained by solving:
                \begin{alignat*}{3}[left=\empheqlbrace]
                -\Delta w&=0, &\qquad&{\text{in }}\Omega,
                \\
                w &= 1, &\qquad&{\text{on }}\Gamma_\mathrm{mv} \cup \Gamma_\mathrm{apex},
                \\
                w &= 0, &\qquad&{\text{on }}\Gamma_\mathrm{av},
                \\
                \nabla w \cdot \mathbf{n}&=0, &\qquad&{\text{on }}\partial \Omega
                \setminus(\Gamma_\mathrm{av} \cup \Gamma_\mathrm{mv} \cup \Gamma_\mathrm{apex}).
                \end{alignat*}

                \item[Piersanti et al. approach \cite{piersanti2021modeling}:] for each point in $\Omega$, a unique normal direction $\boldsymbol{k}$ is selected among the gradient of several normal directions $\boldsymbol{k}=\nabla \psi_\mathrm{i}$ ($\mathrm{i=ab,v,r}$), where $\psi_\mathrm{i}$ are obtained by solving the following \ac{LD} problem
                \begin{equation}[left=\empheqlbrace]
                \label{atrial}
                \begin{alignedat}{3}
                -\Delta \psi_\mathrm{i}&=0, &\qquad&{\text{in }}\Omega,
                \\
                \psi_\mathrm{i} &= \chi_\mathrm{a}, &\qquad&{\text{on }}\Gamma_\mathrm{a},
                \\
                \psi_\mathrm{i} &= \chi_\mathrm{b}, &\qquad&{\text{on }}\Gamma_\mathrm{b},
                \\
                \nabla \psi_{i} \cdot \mathbf{n}&=0, &\qquad&{\text{on }}\partial \Omega
                \setminus(\Gamma_\mathrm{a} \cup \Gamma_\mathrm{b}).
                \end{alignedat}
                \end{equation}
                Please refer to \cite{piersanti2021modeling} for further details about the selection procedure for $\boldsymbol{k}$ and the specific choices of $\chi_\mathrm{a}$, $\chi_\mathrm{b}$, $\Gamma_\mathrm{a}$ and $\Gamma_\mathrm{b}$ in problem \eqref{normal} made for $\psi_\mathrm{i}$ ($i=\mathrm{ab},\mathrm{v},\mathrm{r}$).
        \end{description}

        The \ac{BT} approach is used in the ventricular and spherical slab geometry types, see also Figure~\ref{fig:laplace}(a). The \ac{BT} and \ac{RL} approaches can be adopted in the based ventricular geometry (by setting either \verb|Algorithm type| equal to \verb|BT| or \verb|RL| in the parameter file, respectively), while the Doste approach is used in the complete ventricular geometry, see also Figure~\ref{fig:laplace}(b). Finally, the Piersanti approach is employed for the atrial geometry, see also Figure~\ref{fig:laplace}(c).

        \item[4. Local coordinate system:] For each point of the domain an orthonormal local coordinate axial system is defined by
        $\widehat{\boldsymbol{e}}_t$, $\widehat{\boldsymbol{e}}_n$ and the unit longitudinal direction $\widehat{\boldsymbol{e}}_l$ (orthogonal to the previous ones), as shown in Figure~\ref{fig:directions}:
        \begin{equation}
        Q=\left[\widehat{\boldsymbol{e}}_l, \widehat{\boldsymbol{e}}_n,
        \widehat{\boldsymbol{e}}_t\right]=
        \left\{
        \begin{aligned}
        \widehat{\boldsymbol{e}}_t &=
        \frac{\nabla \phi}{\Vert \nabla \phi \Vert},
        \\
        \widehat{\boldsymbol{e}}_n &= \frac{\boldsymbol{k} - (\boldsymbol{k} \cdot
                \widehat{\boldsymbol{e}}_t )\widehat{\boldsymbol{e}}_t}{\Vert
                \boldsymbol{k} - (\boldsymbol{k} \cdot \widehat{\boldsymbol{e}}_t
                )\widehat{\boldsymbol{e}}_t \Vert},
        \\
        \widehat{\boldsymbol{e}}_l &= \widehat{\boldsymbol{e}}_n \times
        \widehat{\boldsymbol{e}}_t.
        \end{aligned}
        \right.
        \end{equation}
        \item[5. Axis rotation:] The reference frame is rotated with the purpose of defining the myofibers orientation: $\boldsymbol f$ the fiber
        direction, $\boldsymbol n$ the sheet-normal direction and $\boldsymbol s$ the sheet direction. Specifically, $\widehat{\boldsymbol{e}}_l$ rotates
        counter-clockwise around $\widehat{\boldsymbol{e}}_t$ by the helical angle $\alpha$, whereas the transmural direction $\widehat{\boldsymbol{e}}_t$ is rotated counter-clockwise around $\widehat{\boldsymbol{e}}_l$ by the sheetlet angle $\beta$, see Figure~\ref{fig:directions}:
        $$
        \left[\widehat{\boldsymbol{e}}_l, \widehat{\boldsymbol{e}}_n,
        \widehat{\boldsymbol{e}}_t\right] \longrightarrow
        \left[\boldsymbol f, \boldsymbol n, \boldsymbol s\right],
        $$
        The rotation angles follow the linear relationships:
        $$
        \alpha(\phi) = \alpha_\mathrm{endo}(1-\phi)+\alpha_\mathrm{epi}\phi,
        \qquad
        \beta(\phi) = \beta_\mathrm{endo}(1-\phi)+\beta_\mathrm{epi}\phi,
        $$
        where $\alpha_\mathrm{endo}$, $\alpha_\mathrm{epi}$, $\beta_\mathrm{endo}$, $\beta_\mathrm{epi}$ are suitable helical and sheetlet rotation angles on the epicardium and endocardium (specifying in the parameter file \verb|alpha epi|, \verb|alpha endo|, \verb|beta epi|, \verb|beta endo|). Moreover, for the complete ventricular geometry it is possible to set specific fiber and sheet angle rotations in the outflow tract (OT) region (\textit{i.e.} around the aortic valve ring) by specifying \verb|alpha epi OT|, \verb|alpha endo OT|, \verb|beta epi OT|, \verb|beta endo OT|). Finally, for the atrial geometry type, no transmural variation in the myofibers direction is prescribed and the three unit directions correspond to the final myofibers directions $[\widehat{\boldsymbol{e}}_l, \widehat{\boldsymbol{e}}_n, \widehat{\boldsymbol{e}}_t]
        =[\boldsymbol f, \boldsymbol n, \boldsymbol s]$.
\end{description}

\begin{figure}
	\centering
	\includegraphics[width=0.9\textwidth]{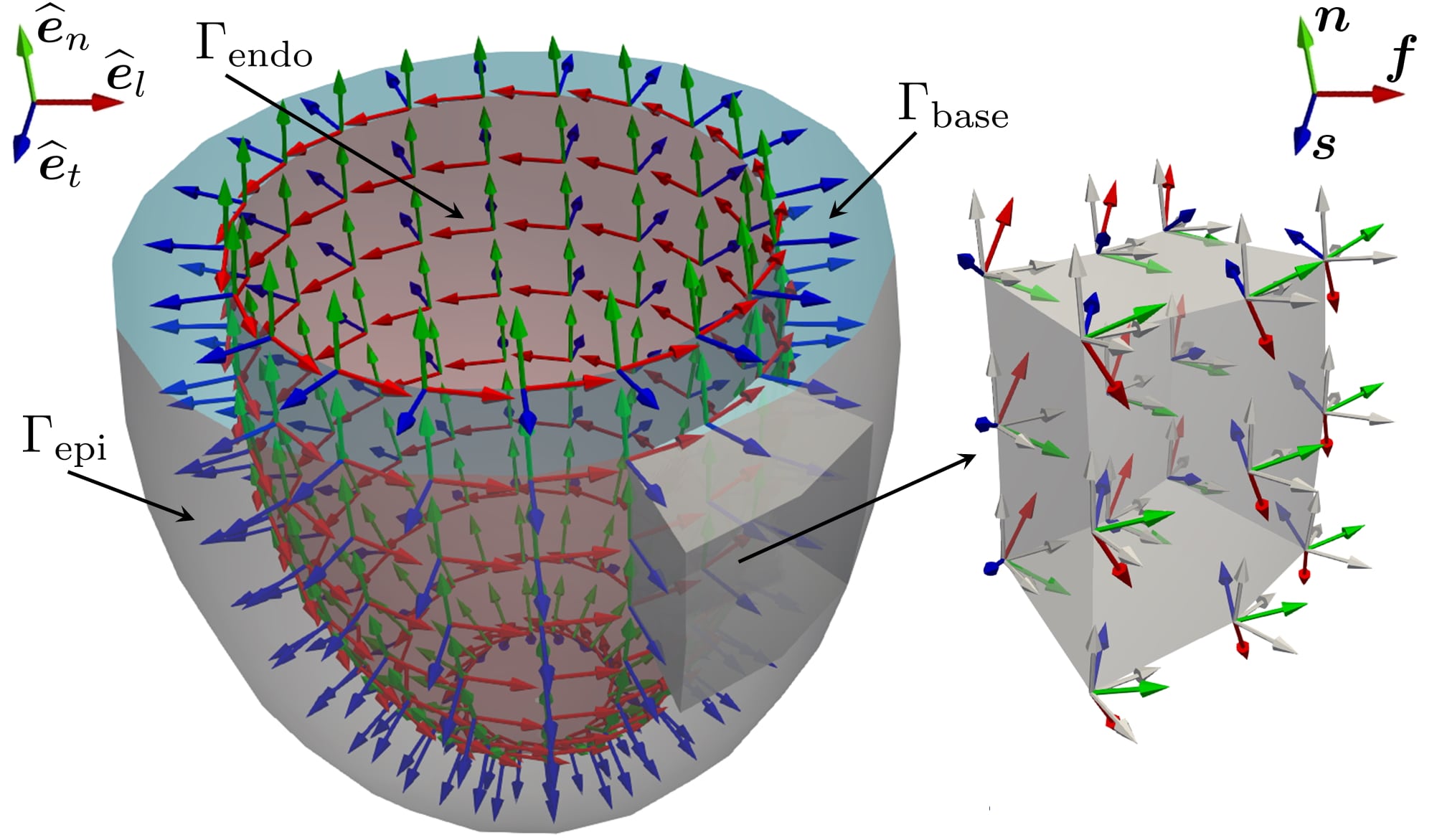}
	\caption{Representation of the local coordinate system employed by a \ac{LDRBM} for an idealized ventricular domain. Only directions on the endocardium $\Gamma_\mathrm{endo}$ are represented. In blue: unit transmural direction, $\widehat{\boldsymbol{e}}_t$. In green: unit normal direction, $\widehat{\boldsymbol{e}}_n$. In red: unit longitudinal direction, $\widehat{\boldsymbol{e}}_l$. Right: zoom on a slab of the left ventricular myocardium  showing the three final myofibers orientations $\boldsymbol f$, $\boldsymbol s$ and $\boldsymbol n$.}
	\label{fig:directions}
\end{figure}

In order to represent the fiber architecture, \acp{LDRBM} use the gradient of specific intra-chamber distances, by means of harmonic problems, combined with a precise definition of boundary sections where boundary conditions are prescribed. This strategy makes the fibers less open to subjective variability. On the other hand, the myofiber orientations could be adapted to a patient-specific setting by simply changing the parameters involved in \acp{LDRBM} (\textit{e.g.} the helical and sheetlet angles $\alpha$ and $\beta$). Therefore, unlike other \acp{RBM} requiring manual or semi-automatic interventions, \acp{LDRBM} can be easily applied to any arbitrary patient-specific geometry~\cite{piersanti2021modeling}.

\subsection*{Comparison to existing software}\label{sec:comparison}
Several packages have been developed and are available in the framework of cardiac fibers generation.

\texttt{Meshtools}\footnote{\url{https://bitbucket.org/aneic/meshtool/src/master}} \cite{meshtools} is a comand-line tool designed to automate image-based mesh generation and manipulate tasks in cardiac modeling workflows, such as operations on label fields and/or geometric features; it integrates seamlessly with the \texttt{openCARP}\footnote{\url{https://opencarp.org/}} ecosystem \cite{openCARP}; the algorithms supported are only for left ventricular geometries and of \ac{BT} type.
\texttt{KIT-IBT-LDRB\_Fibers}\footnote{\url{https://github.com/KIT-IBT/LDRB_Fibers}} is a \texttt{MATLAB} tool for generating left and bi-ventricular fibers; the original \ac{BT} algorithm was adapted to eliminate a discontinuity in the fiber field in correspondence of the free walls and to yield a fiber rotation that is directly proportional to the transmural Laplace solution (approximately linear across the wall) \cite{kovacheva2021causes}.
\texttt{CARDIO SUITE for GIMIAS}\footnote{\url{http://www.gimias.org/index.php/clinical-prototypes/cardiosuite}} includes tools for patient-specific modelling that allow to generate the \ac{FE} meshes required for the simulations and to build additional structures such as fiber orientation \cite{larrabide2009gimias}; at the time of writing and to the best of our knowledge, the latest version was released in 2016. \texttt{SimCardio}\footnote{\url{http://simvascular.github.io/index.html}} is advertised as the only fully open-source software package providing a complete pipeline from medical image data segmentation to patient specific blood flow simulation and analysis; its module \texttt{svFSI} supports specifying distributed fiber and sheet direction generated by \ac{BT}-like rule-based algorithms.

Poisson interpolation algorithms \cite{wong2014generating} have inspired the implementation of fiber generation packages, despite being generally more computationaly demanding than, \textit{e.g.}, \ac{RL} or \ac{BT} algorithms \cite{rossi2014thermodynamically}. This class of methods has been implemented in \texttt{Cardiac Chaste}\footnote{\url{https://www.cs.ox.ac.uk/chaste/cardiac_index.html}}, which
supports automatic generation of mathematical model for fiber orientation associated with both idealized and anatomically-based geometry meshes \cite{cooper2020chaste}, and in \texttt{BeatIt}\footnote{\url{https://github.com/rossisimone/beatit}}, which is to our knowledge the only publicly available software which natively supports the generation of fiber architectures also for atrial geometries \cite{rossi2018muscle}.

Compared to the software described above, the strengths of \lifex{} reside in its user-friendly interface and in its generality, by supporting either idealized and realistic, (left) ventricular and atrial geometries and for each of them the user can selected one of the different state-of-the-art algorithms described in Section~\nameref{sec:methods}. Finally, \lifex{} offers a seamless integration with many other core models (specifically cardiac electrophysiology, mechanics and electromechanics) for modeling the cardiac function from single-chamber to whole-heart simulations which will be targeted by future releases.

\section*{Implementation}
In this section we introduce technical specifications of \lifex{} as well as a thorough documentation of the user interface exposed. The users will be guided from downloading it to running a full simulation of the algorithms presented in Section~\nameref{sec:methods}.

\subsection*{Technical specifications}\label{sec:techspec}
Here we specify the package content, copyright and licensing information and software and hardware specifications required by the present \lifex{} release.

\subsubsection*{Package content}
The fiber generation module of \lifex{} is delivered in binary form as an
\texttt{AppImage}\footnote{\url{https://appimage.org/}} executable.

This provides a universal package for \texttt{x86-64 Linux} operating systems, without the need to deliver different distribution-specific versions. From the user's perspective, this implies an effortless \textit{download-then-run} process, without having to manually take care of installing the proper system dependencies required.

Once the source code will be made publicly accessible, a standard \textit{build-from-source} procedure with automatic installers will be available to make the dependencies setup tailored to the specific hardware of \ac{HPC} facilities or cloud platforms.

\subsubsection*{License and third-party software}
This work is copyrighted by the \lifex{} authors and licensed under the Creative Commons Attribution Non-Commercial No-Derivatives 4.0 International License\footnote{\url{http://creativecommons.org/licenses/by-nc-nd/4.0/}}.

\lifex{} makes use of third-party libraries.
Please note that such libraries are copyrighted by their
respective authors (independent of the \lifex{}
authors) and are covered by various permissive licenses.

The third-party software bundled with (in binary form),
required by, copied, modified or explicitly used in
\lifex{} include the following packages:
\begin{description}
\item[\texttt{Boost}\footnote{\url{https://www.boost.org/}}:] its modules \texttt{Filesystem} and \texttt{Math} are used for manipulating files/directories and for advanced mathematical functions and interpolators, respectively;
\item[\texttt{deal.II}\footnote{\url{https://www.dealii.org/}}:] it provides support to mesh handling, assembling and solving Finite Element problems (with a main  support to third-party libraries as \texttt{PETSc}\footnote{\url{https://www.mcs.anl.gov/petsc/}} and \texttt{Trilinos}\footnote{\url{https://trilinos.github.io/}} for linear algebra data structures and solvers) and to input/output functionalities;
\item[\texttt{VTK}\footnote{\url{https://vtk.org/}}:] it is used for importing external surface or volume input data and coefficients appearing in the mathematical formulation.
\end{description}
Some of the packages listed above, as stated by their respective authors, rely on additional third-party dependencies that may also be bundled (in binary form) with
\lifex{}, although not used directly.
These dependencies include:
\texttt{ADOL-C}\footnote{\url{https://github.com/coin-or/ADOL-C}},
\texttt{ARPACK-NG}\footnote{\url{https://github.com/opencollab/arpack-ng}},
\texttt{BLACS}\footnote{\url{https://www.netlib.org/blacs/}},
\texttt{Eigen}\footnote{\url{https://eigen.tuxfamily.org/}},
\texttt{FFTW}\footnote{\url{https://www.fftw.org/}},
\texttt{GLPK}\footnote{\url{https://www.gnu.org/software/glpk/}},
\texttt{HDF5}\footnote{\url{https://www.hdfgroup.org/solutions/hdf5/}},
\texttt{HYPRE}\footnote{\url{https://www.llnl.gov/casc/hypre/}},
\texttt{METIS}\footnote{\url{http://glaros.dtc.umn.edu/gkhome/metis/metis/overview}},
\texttt{MUMPS}\footnote{\url{http://mumps.enseeiht.fr/index.php?page=home}},
\texttt{NetCDF}\footnote{\url{https://www.unidata.ucar.edu/software/netcdf/}},
\texttt{OpenBLAS}\footnote{\url{https://www.openblas.net/}},
\texttt{ParMETIS}\footnote{\url{http://glaros.dtc.umn.edu/gkhome/metis/parmetis/overview}},
\texttt{ScaLAPACK}\footnote{\url{https://www.netlib.org/scalapack/}},
\texttt{Scotch}\footnote{\url{https://gitlab.inria.fr/scotch/scotch}},
\texttt{SuiteSparse}\footnote{\url{https://people.engr.tamu.edu/davis/suitesparse.html}},
\texttt{SuperLU}\footnote{\url{https://portal.nersc.gov/project/sparse/superlu/}},
\texttt{oneTBB}\footnote{\url{https://oneapi-src.github.io/oneTBB/}},
\texttt{p4est}\footnote{\url{https://www.p4est.org/}}.

The libraries listed above are all free software and, as such, they place
few restrictions on their use. However, different terms may hold.
Please refer to the content of the folder \verb|doc/licenses/| for more information on license and copyright statements for these packages.

Finally, an \texttt{MPI} installation (such as \texttt{OpenMPI}\footnote{\url{https://www.open-mpi.org/}} or \texttt{MPICH}\footnote{\url{https://www.mpich.org/}}) may also be required to successfully run \lifex{} executables in parallel.

\subsubsection*{Software and hardware requirements}
As an \texttt{AppImage}, this \lifex{} release has been built on \texttt{Debian Buster} (the current \texttt{oldstable} version)\footnote{\url{https://www.debian.org/releases/}} following the \textit{``Build on old systems, run on newer systems''} paradigm\footnote{\url{https://docs.appimage.org/introduction/concepts.html}}.

Therefore, it is expected to run on (virtually) any \textit{recent enough} \texttt{x86-64 Linux} distribution, assuming that \texttt{glibc}\footnote{\url{https://www.gnu.org/software/libc/}} version \texttt{2.28} or higher is installed.

\subsection*{Quick start guide: running a \lifex{} executable}\label{sec:run}
The following steps are required in order to run a \lifex{} executable.

\subsubsection*{Download and install}\label{sec:download}
First, download and extract the \lifex{} release archive available at \url{https://doi.org/10.5281/zenodo.5810268}. Then\footnote{\url{https://docs.appimage.org/introduction/quickstart.html}}\,\footnote{\url{https://docs.appimage.org/user-guide/run-appimages.html}}:
\begin{enumerate}
    \item open a terminal;
    \item move to the directory containing the \\ \verb|lifex_fiber_generation-1.4.0-x86_64.AppImage| file:
\begin{lstlisting}[language=bash]
cd /path/to/lifex/
\end{lstlisting}
    \item make the \texttt{AppImage} executable:
\begin{lstlisting}[language=bash]
chmod +x lifex_fiber_generation-1.4.0-x86_64.AppImage
\end{lstlisting}
    \item \verb|lifex_fiber_generation-1.4.0-x86_64.AppImage| is now ready to be executed:
\begin{lstlisting}[language=bash]
./lifex_fiber_generation-1.4.0-x86_64.AppImage [ARGS]...
\end{lstlisting}
\end{enumerate}

Root permissions are \textbf{not} required. Please note that, in order for the above procedure to succeed, \texttt{AppImage} relies upon the userspace filesystem framework \texttt{FUSE}\footnote{\url{https://www.kernel.org/doc/html/latest/filesystems/fuse.html}} which is assumed to be installed on your system. In case of errors, please try with the following commands:
\begin{lstlisting}[language=bash]
./lifex_fiber_generation-1.4.0-x86_64.AppImage \
  --appimage-extract

squashfs-root/usr/bin/lifex_fiber_generation [ARGS]...
\end{lstlisting}
or refer to the \texttt{AppImage} troubleshooting guide\footnote{\url{https://docs.appimage.org/user-guide/troubleshooting/fuse.html}}.

\subsubsection*{Step 0 - Set the parameter file}
Each \lifex{} application or example (including also the fiber generation executable described in Section~\nameref{sec:download}) defines a set of parameters that
are required in order to be run. They involve problem-specific
parameters (such as constitutive relations, geometry, time interval,
boundary conditions, ...) as well as numerical parameters (types of linear/non-linear solvers, tolerances, maximum number of iterations, ...) or output-related options.

In case an application has sub-dependencies (such as a linear solver), also the related parameters are included (typically in a proper subsection).

Every application comes with a set of command line options, which can be printed using the \verb|-h| (or \verb|--help|) flag:
\begin{lstlisting}[language=bash]
./executable_name -h
\end{lstlisting}

The first step before running an executable is to generate the parameter
file(s) containing all the default parameter values. This is done via the
\verb|-g| (or \verb|--generate-params|) flag:
\begin{lstlisting}[language=bash]
./executable_name -g
\end{lstlisting}
that by default generates a parameter file named after the executable, in \verb|.prm| format.

By default, only parameters considered \textit{standard} are printed. The parameter file verbosity can be decreased or increased by passing an optional flag
\verb|minimal| or \verb|full| to the \verb|-g| flag, respectively:
\begin{lstlisting}[language=bash]
./executable_name -g minimal
\end{lstlisting}

The parameter basename to generate can be customized with the \verb|-f| (or \verb|--params-filename|) option:
\begin{lstlisting}[language=bash]
./executable_name -g -f custom_param_file.ext
\end{lstlisting}
Absolute or relative paths can be specified.

At user's option, in order to guarantee a flexible interface to external file processing tools, the parameter file extension \verb|ext| can be chosen among three different interchangeable file formats \verb|prm|, \verb|json| or \verb|xml|, from the most human-readable to the most machine-readable.

As an example, the three parameter files listed in Listings~\ref{lst:prm}, \ref{lst:json} and \ref{lst:xml} are semantically equivalent.

We highlight that, following with the design of the \texttt{ParameterHandler} class of \texttt{deal.II}\footnote{\scriptsize\url{https://www.dealii.org/current/doxygen/deal.II/classParameterHandler.html}}, each parameter is provided with:
\begin{itemize}
    \item a given pattern, specifying the parameter type (\textit{e.g.} boolean, integer, floating-point number, string, list, \dots) and, whenever relevant, a range of admissible values (\textit{pattern description});
    \item a default value, printed in the parameter file upon generation and implicitly assumed if the user omits a custom value;
    \item the \textit{actual} value, possibly overriding the default one;
    \item a documentation string;
    \item a global index.
\end{itemize}

All of these features, combined to a runtime check for the correctness of each parameter, make the code syntactically and semantically robust with respect to possible errors or typos introduced unintentionally.

\begin{lstlisting}[language=prm, label=lst:prm, caption={\texttt{PRM} parameter file format.}]
# Listing of Parameters
# ---------------------
subsection Fiber generation
  subsection Mesh and space discretization

    # Specify whether the input mesh has hexahedral or
    # tetrahedral elements. Available options are: Hex | Tet.
    set Element type          = Hex

    # Number of global mesh refinement steps applied to the initial grid (Hex only).
    set Number of refinements = 0

    subsection File
      # Mesh file.
      set Filename       =

      # Mesh scaling factor: 1e-3 => from [mm] to [m].
      set Scaling factor = 1e-3
    end
  end

  subsection Output
    # Enable/disable output.
    set Enable output = true

    # Output file.
    set Filename      = fibers
  end
end
\end{lstlisting}

\begin{lstlisting}[language=json, label=lst:json, caption={\texttt{JSON} parameter file format.}]
{
  "Fiber_20generation": {
    "Mesh_20and_20space_20discretization": {
      "File": {
        "Filename": {
          "value": "",
          "default_value": "",
          "documentation": "Mesh file.",
          "pattern": "0",
          "pattern_description": "[FileName (Type: input)]"
        },
        "Scaling_20factor": {
          "value": "1",
          "default_value": "1e-3",
          "documentation": "Mesh scaling factor: 1e-3 => from [mm] to [m].",
          "pattern": "1",
          "pattern_description": "[Double 0...MAX_DOUBLE (inclusive)]"
        }
      },
      "Element_20type": {
        "value": "Hex",
        "default_value": "Hex",
        "documentation": "Specify whether the input mesh has hexahedral or tetrahedral elements. Available options are: Hex | Tet.",
        "pattern": "2",
        "pattern_description": "[Selection Hex|Tet ]"
      },
      "Number_20of_20refinements": {
        "value": "0",
        "default_value": "0",
        "documentation": "Number of global mesh refinement steps applied to the initial grid (Hex only).",
        "pattern": "3",
        "pattern_description": "[Integer range 0...2147483647 (inclusive)]"
      }
    },
    "Output": {
      "Enable_20output": {
        "value": "true",
        "default_value": "true",
        "documentation": "Enable\/disable output.",
        "pattern": "4",
        "pattern_description": "[Bool]"
      },
      "Filename": {
        "value": "fibers",
        "default_value": "fibers",
        "documentation": "Output file.",
        "pattern": "5",
        "pattern_description": "[FileName (Type: output)]"
      }
    }
  }
}
\end{lstlisting}

\begin{lstlisting}[language=xml, label=lst:xml, caption={\texttt{XML} parameter file format.}]
<?xml version="1.0" encoding="utf-8"?>
<ParameterHandler>
  <Fiber_20generation>
    <Mesh_20and_20space_20discretization>
      <File>
        <Filename>
          <value/>
          <default_value/>
          <documentation>Mesh file.</documentation>
          <pattern>0</pattern>
          <pattern_description>[FileName (Type: input)]</pattern_description>
        </Filename>
        <Scaling_20factor>
          <value>1</value>
          <default_value>1</default_value>
          <documentation>Mesh scaling factor: 1e-3 =&gt; from [mm] to [m].</documentation>
          <pattern>1</pattern>
          <pattern_description>[Double 0...MAX_DOUBLE (inclusive)]</pattern_description>
        </Scaling_20factor>
      </File>
      <Element_20type>
        <value>Hex</value>
        <default_value>Hex</default_value>
        <documentation>Specify whether the input mesh has hexahedral or tetrahedral elements. Available options are: Hex | Tet.</documentation>
        <pattern>2</pattern>
        <pattern_description>[Selection Hex|Tet ]</pattern_description>
      </Element_20type>
      <Number_20of_20refinements>
        <value>0</value>
        <default_value>0</default_value>
        <documentation>Number of global mesh refinement steps applied to the initial grid (Hex only).</documentation>
        <pattern>3</pattern>
        <pattern_description>[Integer range 0...2147483647 (inclusive)]</pattern_description>
      </Number_20of_20refinements>
    </Mesh_20and_20space_20discretization>
    <Output>
      <Enable_20output>
        <value>true</value>
        <default_value>true</default_value>
        <documentation>Enable/disable output.</documentation>
        <pattern>4</pattern>
        <pattern_description>[Bool]</pattern_description>
      </Enable_20output>
      <Filename>
        <value>fibers</value>
        <default_value>fibers</default_value>
        <documentation>Output file.</documentation>
        <pattern>5</pattern>
        <pattern_description>[FileName (Type: output)]</pattern_description>
      </Filename>
    </Output>
  </Fiber_20generation>
</ParameterHandler>
\end{lstlisting}

Once generated, the user can modify, copy, move or rename the parameter file depending on their needs.

\subsubsection*{Step 1 - Run}
To run an executable, the \verb|-g| flag has simply to be omitted
whereas the \verb|-f| option is used to specify the
parameter file to be \textit{read} (as opposed to \textit{written}, in generation mode), \textit{e.g.}:
\begin{lstlisting}[language=bash]
./executable_name -f custom_param_file.ext [option...]
\end{lstlisting}

If no \verb|-f| flag is provided, a file named \verb|executable_name.prm| is
assumed to be available in the directory where the executable is run from.

The path to the directory where all the app output files will be saved to
can be selected via the \verb|-o| (or \verb|--output-directory|)
flag:
\begin{lstlisting}[language=bash]
./executable_name -o ./results/
\end{lstlisting}

If the specified directory does not already exist, it will be created.
By default, the current working directory is used.

Absolute or relative paths can be specified for both the input parameter file and the output directory.

\paragraph*{Parallel run}
To run an app in parallel, use the \verb|mpirun| or \verb|mpiexec| wrapper commands (which may vary depending on the MPI implementation available on your machine) \textit{e.g.}:
\begin{lstlisting}[language=bash]
mpirun -n <N_PROCS> ./executable_name [option...]
\end{lstlisting}
where \verb|<N_PROCS>| is the desired number of parallel processes to run on.

As a rule of thumb, 10000 to 100000 degrees of freedom per process should
lead to the best performance.

\paragraph*{Dry run and parameter file conversion}
Upon running, a parameter log file is automatically generated in the output
directory, that can be used later to retrieve which parameters had been used
for a specific run.

By default, \verb|log_params.ext| will be used as its filename. This can
be changed via the \verb|-l| (or \verb|--log-file|) flag, \textit{e.g.}:
\begin{lstlisting}[language=bash]
./executable_name -l my_log_file.ext [option...]
\end{lstlisting}

The extension is not mandatory: if unspecified, the same extension as the
input parameter file will be used.

If the \textbf{dry run} option is enabled via the \verb|-d| (or
\verb|--dry-run|) flag, the execution terminates right after the parameter log file generation. This has a two-fold purpose:
\begin{enumerate}
    \item checking the correctness of the parameters being declared and parsed \textit{before} running the actual simulation (if any of the parameters does not match the specified pattern or has a wrong name or has not been declared in a given subsection then a runtime exception is thrown);
    \item \textbf{converting} a parameter file between two different
    formats/extensions. For example, the following command converts
    \verb|input.xml| to\\\verb|output.json|:
\begin{lstlisting}[language=bash]
./executable_name -f input.xml -d -l output.json [option...]
\end{lstlisting}
\end{enumerate}

\section*{Results and Discussion}\label{sec:results}
The \acp{LDRBM} described in Section~\nameref{sec:methods} have been applied
to a set of idealized or realistic test cases, namely ventricular and spherical slabs, based and complete left ventricular and left atrial geometries.

This section presents the results obtained, as well as a possible pipeline for reproducing such test cases, consisting of the following steps:
\begin{enumerate}
    \item setting up input data (\textit{e.g.} generating or importing computational meshes);
    \item setting up the parameter files associated with a given simulation scenario and running the corresponding simulation;
    \item post-processing the solution and visualizing the output.
\end{enumerate}

A mesh sensitivity analysis has been performed and reported in Section~\nameref{sec:mesh_sensitivity}.

\subsection*{Input data}\label{sec:data}
Additional input data (scripts, meshes and parameter files) associated with the guided examples described below can be downloaded from the release archive \url{https://doi.org/10.5281/zenodo.5810268}.

In this \textit{getting started} guide, we provide different ready-to-use meshes, namely
\begin{itemize}
    \item a set of four idealized geometries consisting of a ventricular slab, a spherical slab, an idealized based left ventricle and an idealized left atrium, see Figures~\ref{fig:tags}(a-d);
    \item two realistic geometries composed by a left ventricle and a left atrium, see Figures~\ref{fig:tags}(e-f).
\end{itemize}

The idealized meshes have been generated using the built-in CAD engine of \texttt{gmsh}, an open-source 3D \ac{FE} mesh generator, starting from the corresponding \texttt{gmsh} geometrical models (represented by \verb|*.geo| files, also provided) defined using their boundary representation, where a volume is bounded by a set of surfaces. For details about the geometrical definition of a model geometry we refer to the online documentation of \texttt{gmsh}\footnote{\url{https://gmsh.info/doc/texinfo/gmsh.html}}.

In order to perform the mesh generation, starting from the geometrical files provided in this tutorial, the following command can be run in a terminal:
\begin{lstlisting}[language=bash]
gmsh geometry.geo -clscale s -o mesh.msh -save
\end{lstlisting}
where \verb|geometry.geo| is the geometrical file model, \verb|mesh.msh| is the output mesh file, which will be provided as an input to a \lifex{} app, and \verb|s| $\in (0,1]$ is the mesh element size factor. To produce a coarser (finer) mesh the \verb|clscale| factor can be reduced (increased).

The realistic left ventricle and left atrium have been produced starting from the open-source meshes adopted in \cite{fastl2018personalized} (for the left atrium\footnote{\url{https://doi.org/10.18742/RDM01-289}}) and in \cite{strocchi2020publicly} (for the left ventricle\footnote{\url{https://doi.org/10.5281/zenodo.3890034}}) and using the Vascular Modelling Toolkit (\texttt{vmtk}) software \cite{antiga2008vascular} along with the semi-automatic meshing tools\footnote{\url{https://github.com/marco-fedele/vmtk}} recently proposed in \cite{fedele2021polygonal}.

All the characteristic informations of the ready-to-use meshes, above described, are reported in~Table~\ref{tab:mesh_info}.

\begin{table}
    \centering
    \begin{tabular}{lccccrrc}
        Geometry & Type & $\text{h}_{\text{min}}\,\text{[mm]}$ & $\text{h}_{\text{avg}}\,\text{[mm]}$ & $\text{h}_{\text{max}}\,\text{[mm]}$ & \#elements & \#vertices & Quality \\\hline
        Ventricular slab & Hex & 4.74  & 5.95 & 6.73 & 988 & 1400 & 1.69 \\
        Ventricular slab & Tet & 3.10 & 6.76 & 8.71 & 2439 & 762 & 2.91 \\
        Spherical slab & Tet & 1.96 & 3.46 & 4.70 & 18455 & 4924 & 2.46 \\
        Idealized left atrium & Tet & 2.78 & 3.43 & 4.15 & 10039 & 3387 & 2.36 \\
        Idealized left ventricle & Tet & 1.71 & 3.53 & 4.93 & 74488 & 16567 & 3.27 \\
        Realistic left ventricle & Tet & 0.51& 1.17 & 1.67 & 1885227 & 330703 & 2.80 \\
        Realistic left atrium & Tet & 0.15 & 1.00 &  1.76 & 112994 & 33297 & 4.23 \\
    \end{tabular}
    \caption{Mesh information regarding the ready-to-use meshes presented in this work. In particular the mesh quality is computed selecting the edge ratio option in the \texttt{ParaView} mesh quality filter.}
    \label{tab:mesh_info}
\end{table}

\subsection*{Generating fibers}\label{sec:fiber_generation}
Parameter files for fiber generation are characterized by a common section named \verb|Mesh and space discretization|. Select \verb|Element type = Tet| for tetrahedral meshes or \verb|Element type = Hex| for hexahedral meshes. Finally, specify in \verb|FE space degree| the degree of the (piecewise continuous) polynomial FE space used to solve the \ac{LD} problems described above. Finally, we remark that \lifex{} internally treats all physical quantities as if they are provided in \ac{SI}: therefore, a \verb|Scaling factor| can be set in order to convert the input mesh from a given unit of measurement (\textit{e.g.} if the mesh coordinates are provided in millimeters then \verb|Scaling factor| must be set equal to \verb|1e-3|).

The \verb|Geometry type| parameters enables to specify the kind of geometry provided in input, in order to apply the proper \ac{LDRBM} algorithm among the ones described above. Once specified, parameters related to the specific algorithm and geometry will be parsed from a subsection named after the value of \verb|Geometry type|.

All parameters missing from the parameter file will take their default value, which is hard-coded.

\begin{lstlisting}[language=prm]
subsection Mesh and space discretization
  set Element type    = Tet
  set FE space degree = 1
  set Geometry type   = Slab
    
  subsection File
    set Filename       = /path/to/mesh/slab.msh
    set Scaling factor = 1e-3
  end
end
\end{lstlisting}

\subsubsection*{Slab fibers}\label{sec:slab}
Specify \verb|Geometry type = Slab| to prescribe fibers in slab geometries and the path of the input mesh file in \verb|Filename|.

\paragraph*{Ventricular slab}\label{sec:ventricle_slab}
For a ventricular slab geometry \verb|Sphere slab = false| must be set. Specify the label for the top (\verb|Tags base up|) and bottom (\verb|Tags base down|) surfaces of the slab, as well as the epicardium (\verb|Tags epi|) and endocardium (\verb|Tags endo|) ones. Finally, prescribe the helical and sheetlet rotation angles at the epicardium and endocardium using the corresponding \verb|alpha epi|, \verb|alpha endo|, \verb|beta epi|, \verb|beta endo| parameters.

\begin{lstlisting}[language=prm]
subsection Slab
  set Sphere slab = false

  set Tags base up   =  50
  set Tags base down =  60
  set Tags epi       =  10
  set Tags endo      =  20
  set alpha epi      = -60
  set alpha endo     =  60
  set beta epi       =  45
  set beta endo      = -45
end
\end{lstlisting}

\paragraph*{Spherical slab}\label{sec:spherical_slab}
For a spherical slab geometry set \verb|Sphere slab = true|. Specify the epicardial coordinates $(x,y,z)$ of the north (\verb|North pole|) and south (\verb|South pole|) poles of the sphere of the slab, and the labels of the endocardium (\verb|Tags endo|) and epicardium (\verb|Tags epi|). Finally, prescribe the helical and sheetlet rotation angles at the epicardium and endocardium in \verb|alpha epi|, \verb|alpha endo|, \verb|beta epi|, \verb|beta endo|. A fiber architecture for the sphere slab with radial fiber $ \boldsymbol{f} $ can be prescribed by setting in the parameter file \verb|Sphere with radial fibers = true|. This consists of exchanging the sheet direction $\boldsymbol{s}$ with the fiber direction $\boldsymbol{f}$. Instead, a tangential (to the epicardial and endocardial surfaces) fiber field $\boldsymbol{f}$ is assigned when \verb|Sphere with radial fibers = false|.
\begin{lstlisting}[language=prm]
subsection Slab
  set Sphere slab               = true
  set Sphere with radial fibers = true

  set North pole = 0 0 0.025
  set South pole = 0 0 -0.025

  set Tags epi  = 10
  set Tags endo = 20

  set alpha epi  = 0
  set alpha endo = 0
  set beta epi   = 0
  set beta endo  = 0
end
\end{lstlisting}

\subsubsection*{Left ventricular fibers}\label{sec:ventricle}
Specify \verb|Geometry type = Left ventricle| to prescribe fibers in a based left ventricular geometry, or \verb|Geometry type = Left ventricle complete| to prescribe fibers in complete left ventricular geometry. Other mesh-related parameters have the same meaning as described above.

\begin{lstlisting}[language=prm]
subsection Mesh and space discretization
  set Element type    = Tet
  set FE space degree = 1
  set Geometry type   = Left ventricle complete

  subsection File
    set Filename       = /path/to/mesh/ventricle.msh
    set Scaling factor = 1e-3
  end
end
\end{lstlisting}

\paragraph*{Based left ventricle}\label{sec:based_ventricle}
Specify the labels for the basal plane (\verb|Tags base|), the epicardium (\verb|Tags epi|) and endocardium (\verb|Tags endo|) of the ventricle. Select \verb|RL| or \verb|BT| approach in \verb|Algorithm type|. Prescribe the helical and sheetlet rotation angles at the epicardium and endocardium in \verb|alpha epi|, \verb|alpha endo|, \verb|beta epi|, \verb|beta endo|. Finally, for the \verb|RL| approach specify the outward normal vector to the basal plane in \verb|Normal to base|, while for the \verb|BT| approach prescribe the apex epicardial coordinates $(x,y,z)$ (\verb|Apex|) of the ventricle.

\begin{lstlisting}[language=prm]
subsection Left ventricle
  set Tags base = 50
  set Tags epi  = 10
  set Tags endo = 20

  set Algorithm type = BT

  set alpha epi  = -60
  set alpha endo =  60
  set beta epi   =  20
  set beta endo  = -20

  subsection RL
    set Normal to base = 0 0 1
  end

  subsection BT
    set Apex = 0 0 0.0601846
  end
end
\end{lstlisting}

Selecting the \verb|RL| approach as \verb|Algorithm type|, this setup can be also exploited in bi-ventricular geometries (not included in the example meshes).
In this case, the mesh must have two additional surface labels in the right ventricular endocardium:
one delimiting the part facing to the septum (\textit{e.g.} 15) and the other for the remaining part (\textit{e.g.} 25).
In this way, it is sufficient to set \verb|Tags endo| as the two endocardial labels (excluding the right septum) and \verb|Tags epi| as the epicardial label and the right endocardial septum label.
Further detail on this approach can be found in~\cite{piersanti2021modeling}.

\begin{lstlisting}[language=prm]
subsection Left ventricle
  ...
  set Tags epi  = 10, 15
  set Tags endo = 20, 25

  set Algorithm type = RL
  ...
end
\end{lstlisting}

\paragraph*{Complete left ventricle}\label{sec:complete_ventricle}
Specify the labels for the mitral (\verb|Tags MV|) and aortic (\verb|Tags AV|) valve rings, the epicardium (\verb|Tags epi|) and endocardium (\verb|Tags endo|) of the ventricle. Prescribe the apex epicardial coordinates $(x,y,z)$ in \verb|Apex|. Finally, define the helical and sheetlet rotation angles at the epicardium and endocardium in \verb|alpha epi|, \verb|alpha endo|, \verb|beta epi|, \verb|beta endo|. A specific helical and sheetlet rotation angles around the outflow tract of the left ventricle (\textit{i.e.} the mitral valve ring) can be specified by setting \verb|alpha epi OT|, \verb|alpha endo OT|, \verb|beta epi OT|, \verb|beta endo OT|.
\begin{lstlisting}[language=prm]
subsection Left ventricle complete
  set Tags MV   = 50
  set Tags AV   = 60
  set Tags epi  = 10
  set Tags endo = 20

  set Apex = 0.0692 0.0710 0.3522

  set alpha epi  = -60
  set alpha endo =  60
  set beta epi   =  20
  set beta endo  = -20

  set alpha epi OT  = 0
  set alpha endo OT = 90
  set beta epi OT   = 0
  set beta endo OT  = 0
end
\end{lstlisting}
\subsubsection*{Left atrial fibers}\label{sec:left_atrium}
Specify \verb|Geometry type = Left atrium| to prescribe fibers in a left atrial geometry. Other mesh-related parameters have the same meaning as described above.

\begin{lstlisting}[language=prm]
subsection Mesh and space discretization
  set Element type    = Tet
  set FE space degree = 1
  set Geometry type   = Left atrium

  subsection File
    set Filename       = /path/to/mesh/atrium.msh
    set Scaling factor = 1e-3
  end
end
\end{lstlisting}

\paragraph*{Idealized left atrium}\label{sec:left_atrium_hollow}
Select \verb|Appendage = false| to prescribe fibers in the hollow sphere geometry. Specify the labels for the mitral valve ring (\verb|Tags MV|), the right (\verb|Tags RPV|) and left (\verb|Tags LPV|) pulmonary veins rings, the epicardium (\verb|Tags epi|) and endocardium (\verb|Tags endo|) of the idealized atrium.
Finally, select the dimension of the atrial bundles: \verb|Tau bundle MV| for the mitral valve bundle; \verb|Tau bundle LPV| and \verb|Tau bundle RPV| for the left and right pulmonary valves rings bundle.
\begin{lstlisting}[language=prm]
subsection Left atrium
  set Appendage = false

  set Tags epi  = 30
  set Tags endo = 10
  set Tags RPV  = 20
  set Tags LPV  = 50
  set Tags MV   = 40

  set Tau bundle MV  = 0.65
  set Tau bundle LPV = 0.85
  set Tau bundle RPV = 0.15
end
\end{lstlisting}

\paragraph*{Realistic left atrium}\label{sec:left_atrium_realistic}
Select \verb|Appendage = true| to prescribe fibers in a realistic left atrial geometry. Specify the labels for the mitral valve ring (\verb|Tags MV|), the right (\verb|Tags RPV|) and left (\verb|Tags LPV|) pulmonary veins rings, the epicardium (\verb|Tags epi|) and endocardium (\verb|Tags endo|) of the idealized atrium. Prescribe in \verb|Apex| the epicardial coordinates $(x,y,z)$ for the apex of the atrial appendage. Finally, select the dimension of the atrial bundles: for the mitral valve bundle \verb|Tau bundle MV|; for the left and right pulmonary valves rings bundle \verb|Tau bundle LPV| and \verb|Tau bundle RPV|, respectively.
\begin{lstlisting}[language=prm]
subsection Left atrium
  set Appendage = true

  set Apex = 83.868 16.369 45.989

  set Tags epi  = 30
  set Tags endo = 10
  set Tags RPV  = 20
  set Tags LPV  = 50
  set Tags MV   = 40

  set Tau bundle MV  = 0.60
  set Tau bundle LPV = 0.90
  set Tau bundle RPV = 0.10
end
\end{lstlisting}

\subsection*{Output and visualization}\label{sec:output}
\begin{figure}
	\centering
	\includegraphics[width=0.95\textwidth]{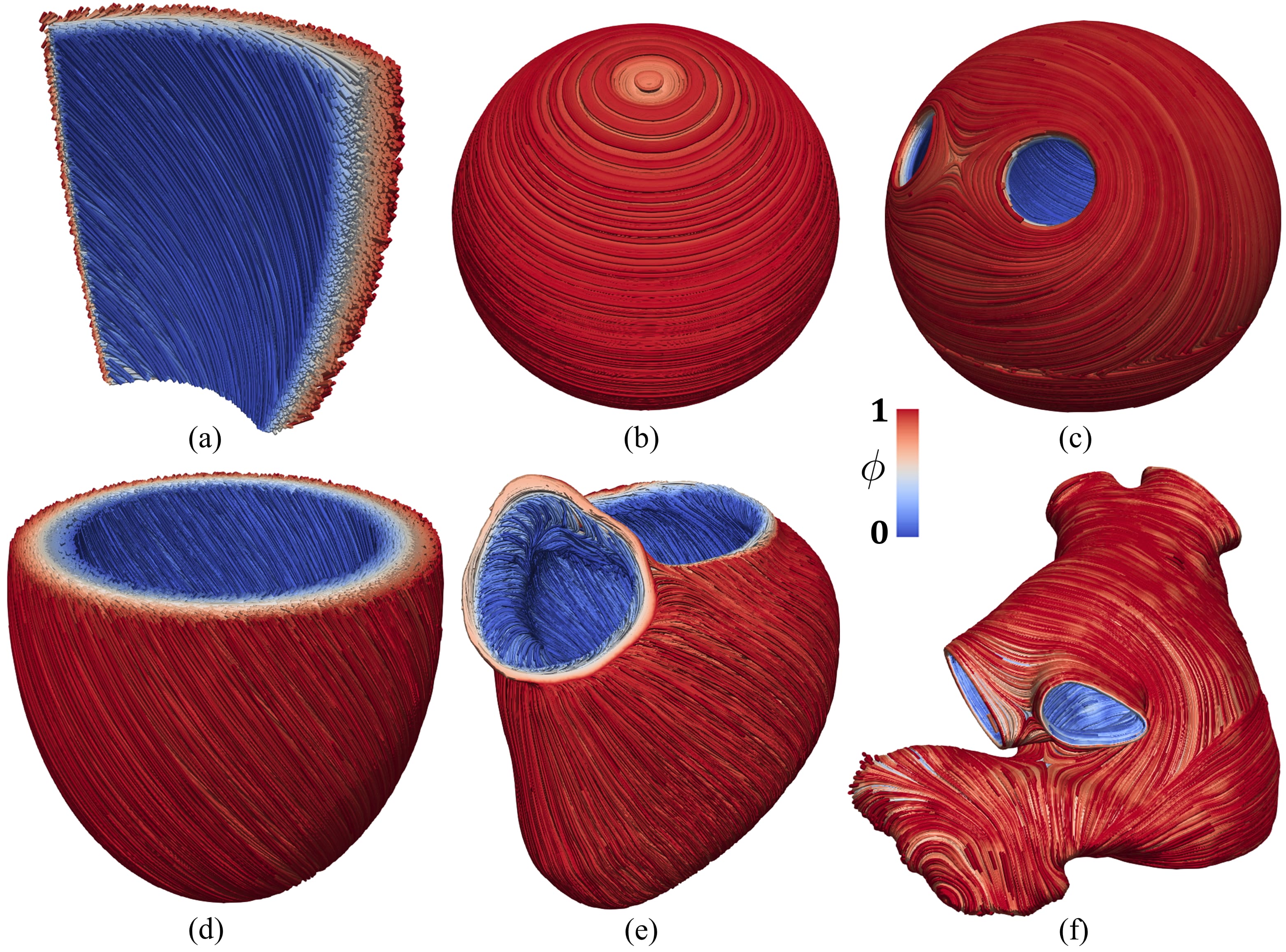}
	\caption{Fiber field $\boldsymbol{f}$ visualized as streamlines. (a) Ventricular slab. (b) Spherical slab with circumferential fibers. (c) Idealized left atrium. (d) Idealized based left ventricle. (e) Realistic complete left ventricle. (f) Realistic left atrium.}
	\label{fig:fibers}
\end{figure}

To enable the output select \verb|Enable output = true| and specify the corresponding output filename in \verb|Filename|. This will produce a \texttt{XDMF} schema file named \verb|output_filename.xdmf| (wrapped around a same-named \texttt{HDF5} output file \verb|output_filename.h5|) that can be visualized in \texttt{ParaView}\footnote{\url{https://www.paraview.org}}, an open-source multi-platform data analysis and visualization application. Specifically, the \texttt{streamtracer} and the \texttt{tube} filters of \texttt{ParaView} can be applied in sequence to visualize the fiber fields, such as the one shown in Figure~\ref{fig:fibers}.

\begin{lstlisting}[language=prm]
subsection Output
  set Enable output = true
  set Filename      = fibers
end
\end{lstlisting}

Moreover, the \texttt{HDF5} file format guarantees that the output can easily be further post-processed, not only for visualization purposes but rather to be fed as an input to more sophisticated computational pipelines.

\subsection*{Mesh sensitivity and validation}\label{sec:mesh_sensitivity}
The robustness of the algorithms presented above is confirmed by performing a mesh sensitivity analysis.
Specifically, we consider the ventricular slab geometry shown in Figure~\ref{fig:tags}(a) and two related sets of discretization with hexahedral and with tetrahedral elements, respectively.
For each of the two element types, we run the fiber generation algorithm (see Section~\nameref{sec:ventricle_slab}) for four decreasing mesh sizes \(h_1 > h_2 > h_3 > h_4\).

As the fibers field is orthonormalized, the difference between two numerical solutions is only due to the orientation angle. This allows us to define an error estimate as:
\begin{equation*}
  \Delta_\theta^{i} = \left|\arccos\left(\mathbf{f}_0^i \cdot \mathbf{f}_0^4\right)\right|,\ i=1,2,3,
\end{equation*}
where \(\mathbf{f}_0^i\) is the fiber field computed on the mesh with size \(h_i\) and the results on the finest mesh with size \(h_4\) are considered as a reference solution.

\begin{table}
    \centering
    \begin{tabular}{ccrcc}
    i & $h\,\text{[mm]}$ & \#dofs & \(\displaystyle\avg_\Omega \Delta_{\theta}^{i}\,[{}^\circ]\) & \(\displaystyle\max_\Omega \Delta_{\theta}^{i}\,[{}^\circ]\) \\\hline
    1 & 5.95 & 1400 & 0.62 & 6.33 \\
    2 & 3.00 & 9477 & 0.24 & 4.54 \\
    3 & 1.50 & 69377 & 0.08 & 2.40 \\
    4 & 0.75 & 530145 & \textemdash & \textemdash \\
    \end{tabular}
    \caption{Mesh sensitivity analysis for hexahedral elements.}
    \label{tab:mesh_hex}
\end{table}

\begin{table}
    \centering
    \begin{tabular}{ccrcc}
    i & $h\,\text{[mm]}$ & \#dofs & \(\displaystyle\avg_\Omega \Delta_{\theta}^{i}\,[{}^\circ]\) & \(\displaystyle\max_\Omega \Delta_{\theta}^{i}\,[{}^\circ]\) \\\hline
    1 & 6.76 & 762 & 1.35 & 7.85 \\
    2 & 3.53 & 3989 & 0.50 & 5.57 \\
    3 & 1.76 & 25661 & 0.19 & 4.28 \\
    4 & 0.88 & 179338 & \textemdash & \textemdash \\
    \end{tabular}
    \caption{Mesh sensitivity analysis for tetrahedral elements.}
    \label{tab:mesh_tet}
\end{table}

\begin{figure}
	\centering
	\includegraphics[width=0.95\textwidth]{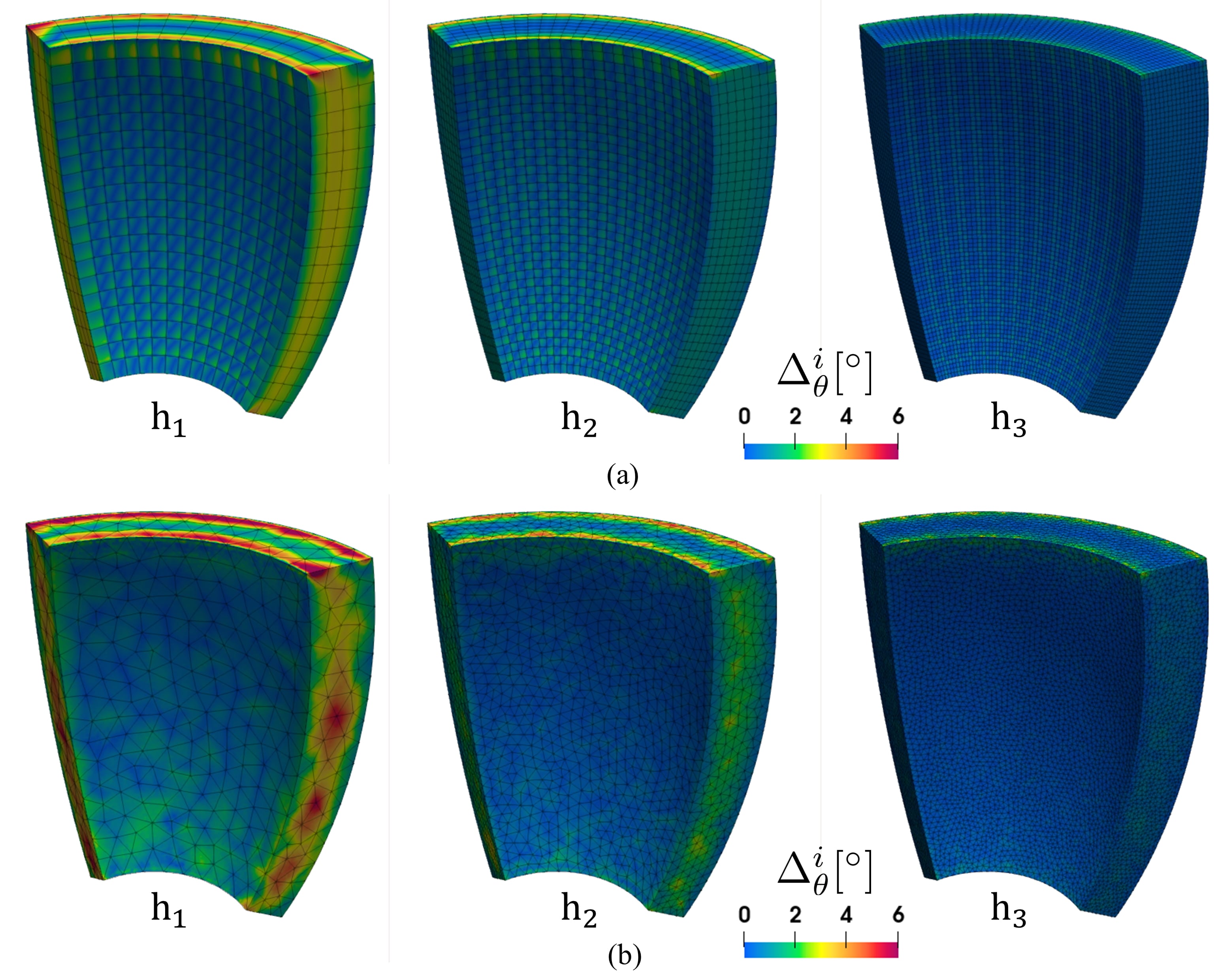}
	\caption{Mesh sensitivity analysis performed on the ventricular slab geometry and three different mesh sizes $h_1,h_2,h_3$. (a) Hexahedral meshes. (b) Tetrahedral meshes.}
	\label{fig:mesh_convergence}
\end{figure}

We show the error distribution in the whole domain in Figure~\ref{fig:mesh_convergence}, while Tables~\ref{tab:mesh_hex} and \ref{tab:mesh_tet} report the average and maximum errors for the hexahedral and the tetrahedral meshes, respectively.
We determine both an average and a maximum error smaller than 8${}^\circ$ even for the coarsest mesh. These values are very small and, in particular, much smaller than the physiological fibers dispersion angle~\cite{guan2021modelling}.

The interest of accurately reconstructing a fiber orientation field consists of providing it as an input to more sophisticated computational models, such as for cardiac electrophysiology and electromechanics. As the dynamics of such physical models demands for a high resolution in both time and space \cite{piersanti2021modeling,piersanti20213d0d,Fedele2022}, the \ac{LDRBM} algorithms are typically run on fine meshes. Therefore, the impact of possible numerical errors due to the mesh size is to be considered negligible, as confirmed by the small errors presented above.

The proposed \acp{LDRBM} have been validated in~\cite{piersanti2021modeling}, where the fiber orientation field computed numerically provided a satisfactory match to histological data.

Furthermore, the anisotropic nature of the fiber orientation strongly influences the electrophysiological, mechanical and electromechanical cardiac function. Therefore, an indirect way to validate the fiber field provided by \acp{LDRBM} is to compute quantitative indices or biomarkers in such kind of simulations.

Matching quantitative indices in a physiological range is only possible when the fiber reconstruction algorithm is accurate enough: this is shown in~\cite{piersanti2021modeling} for the whole-heart electrophysiology, in~\cite{piersanti20213d0d} for the ventricular electromechanics, and in~\cite{Fedele2022} for the whole-heart electromechanics.

\subsection*{Future developments}
As anticipated in Section~\nameref{sec:intro}, the development of \lifex{} has been initially oriented towards a \texttt{heart} module incorporating packages for the simulation of cardiac electrophysiology, mechanics, electromechanics and blood fluid dynamics models.

In the near future, the deployment of \lifex{} will follow two lines:
\begin{itemize}
\item more modules will be successively published in binary form, starting from an advanced solver for cardiac electrophysiology and other solvers for the cardiac function;
\item in the meantime, the source code associated with \lifex{} core and with previous binary releases will be gradually made publicly available under an open-source license.
\end{itemize}

In the long run, also modules unrelated to the cardiac function are expected to be included within \lifex{}.

\section*{Conclusions}\label{sec:conclusions}
\lifex{} is intended to provide the scientific community with an integrated \ac{FE} framework for exploring many physiological and pathological scenarios using \textit{in silico} experiments for the whole-heart cardiac function, boosting both the user and developer experience without sacrificing its computational efficiency and universality.

We believe that this software release provides the scientific community with an invaluable tool for \textit{in silico} scenario analyses of myofibers orientation; such a tool supports either idealized and realistic, (left) ventricular and atrial geometries. It also offers a seamless integration of \acp{LDRBM} into more sophisticated computational pipelines involving other core models -- such as electrophysiology, mechanics and electromechanics -- for the cardiac function, in a wide range of settings covering from single-chamber to whole-heart simulations.

The content of this initial release is published on the official website \url{https://lifex.gitlab.io/}: we encourage users to interact with the \lifex{} development community via the \texttt{issue tracker}\footnote{\url{https://gitlab.com/lifex/lifex.gitlab.io/-/issues}} of our public website repository. Any curiosity, question, bug report or suggestion is welcome!

News and announcements about \lifex{} will be posted to the official website \url{https://lifex.gitlab.io/}. Stay tuned!

\section*{Availability and requirements}
\textbf{Project name}: \lifex{} \\
\textbf{Project home page}: \url{https://lifex.gitlab.io/} \\
\textbf{Operating system(s)}: \texttt{Linux} (\texttt{x86-64})\\
\textbf{Programming language}: \texttt{C++} \\
\textbf{Other requirements}: \texttt{glibc} version \texttt{2.28} or higher \\
\textbf{License}: CC BY-NC-ND 4.0 \\
\textbf{Any restrictions to use by non-academics}: no additional restriction.

\section*{List of abbreviations}
\begin{acronym}
\acro{FE}{Finite Element}
\acro{BT}{Bayer-Trayanova}
\acro{RL}{Rossi-Lassila}
\acro{HPC}{High Performance Computing}
\acro{LD}{Laplace-Dirichlet}
\acro{RBM}{Rule-Based Method}
\acro{LDRBM}{\acl*{LD} \acl*{RBM}}
\acro{SI}{International System of Units}
\end{acronym}

%\begin{backmatter}

\section*{Acknowledgements}\label{sec:funding}
This project has received funding from the European Research Council (ERC) under the European Union's Horizon 2020 research and innovation program (grant agreement No 740132, iHEART - An Integrated Heart Model for the simulation of the cardiac function, P.I. Prof. A. Quarteroni).

\bibliographystyle{bmc-mathphys}
\bibliography{bibliography}

%\end{backmatter}
\end{document}